# A novel method to enhance pneumonia detection via a model-level ensembling of CNN and vision transformer

Sandeep Angara – sandeepangara@gmail.com
Nishith Reddy Mannuru – NishithReddyMannuru@my.unt.edu
Aashrith Mannuru - arm210018@utdallas.edu
Sharath Thirunagaru Sharath.tds@gmail.com

**Abstract:**

Pneumonia remains a leading cause of morbidity and mortality worldwide. Chest X-ray (CXR) imaging is a fundamental diagnostic tool, but traditional analysis relies on time-intensive expert evaluation. Recently, deep learning has shown immense potential for automating pneumonia detection from CXRs. This paper explores applying neural networks to improve CXR-based pneumonia diagnosis. We developed a novel model fusing Convolution Neural networks (CNN) and Vision Transformer networks via model-level ensembling. Our fusion architecture combines a ResNet34 variant and a Multi-Axis Vision Transformer small model. Both base models are initialized with ImageNet pre-trained weights. The output layers are removed, and features are combined using a flattening layer before final classification. Experiments used the Kaggle pediatric pneumonia dataset containing 1,341 normal and 3,875 pneumonia CXR images. We compared our model against standalone ResNet34, Vision Transformer, and Swin Transformer Tiny baseline models using identical training procedures. Extensive data augmentation, Adam optimization, learning rate warmup, and decay were employed. The fusion model achieved a state-of-the-art accuracy of 94.87%, surpassing the baselines. We also attained excellent sensitivity, specificity, kappa score, and positive predictive value. Confusion matrix analysis confirms fewer misclassifications. The ResNet34 and Vision Transformer combination enables jointly learning robust features from CNN's and Transformer paradigms. This model-level ensemble technique effectively integrates their complementary strengths for enhanced pneumonia classification.

1. Introduction

Pneumonia remains a significant global health concern, causing substantial morbidity and mortality across all age groups. According to the World Health Organization (2022) [1], pneumonia is the single largest infectious cause of death in children worldwide, responsible for 14% of all deaths among children under five. This translated to the deaths of 740,180 children under 5 years due to pneumonia in 2019 alone. Pneumonia disproportionately affects populations in low- and middle-income countries, with the highest mortality rates observed in Africa and Asia. The burden of pneumonia in children was most pronounced in Africa and Southeast Asia, accounting for approximately 30% and 39% of total severe pneumonia cases globally [2]. 15 countries in these two areas were responsible for two-thirds of all cases of pneumonia [3]. Beyond mortality, pneumonia also leads to long-term morbidity, such as impaired lung function. Therefore, timely and accurate detection of pneumonia is crucial for prompt treatment and reduction of pneumonia-related deaths and disabilities. Chest X-ray (CXR) imaging has long been a fundamental diagnostic tool for pneumonia and other lung diseases [4]. However, traditional CXR analysis relies extensively on expert human evaluation, which can be time-consuming, resource-intensive, and prone to subjectivity. With rapid advancements in artificial intelligence and deep learning, there has been growing interest in leveraging neural networks to automate and improve pneumonia detection from CXRs

[5]. This research paper explores the application of neural networks for automated pneumonia detection in chest X-rays, investigating the challenges, methodologies, and implications of this emerging technology.

Pneumonia diagnosis traditionally involves the visual inspection and interpretation of chest X-ray images by trained radiologists. This process, while effective, is time-consuming and subject to human error [6]. Studies have found significant inter-observer variability in CXR interpretation among radiologists, underscoring the need for automated systems [7]. Automated pneumonia detection using neural networks aims to streamline this process, reducing diagnostic time and enhancing accuracy. Neural networks, a subset of machine learning algorithms, have shown remarkable capabilities in image analysis tasks, making them promising candidates for pneumonia detection [8].

Deep learning, a subset of neural networks, has achieved remarkable breakthroughs in various domains, particularly image recognition and classification [9]. In medical imaging, neural networks have demonstrated exceptional performance in diagnosing a wide range of conditions, from detecting tumors in MRI scans to identifying diabetic retinopathy in fundus images [10]. These successes have spurred research into their application for pneumonia detection in CXR images.

Applying neural networks for pneumonia detection in CXR images presents unique challenges and considerations. One of the primary challenges is the inherent class imbalance in medical datasets, where healthy cases far outnumber pneumonia cases. Model generalization and robustness across diverse populations and imaging devices also remain critical concerns. Variations in imaging settings, patient positioning, and populations can impact model performance [11]. Ethical considerations, such as transparency in model decision-making and potential bias, demand careful attention. Interpretability of neural network decisions in medical diagnosis, often referred to as "black-box" concerns, raises important questions about the clinical adoption of these models.

The successful integration of neural networks into pneumonia detection could revolutionize the field of radiology by providing radiologists with an auxiliary tool to enhance accuracy and expedite diagnoses. Such advancements could particularly benefit regions with limited access to experienced radiologists. However, it is crucial to validate these models rigorously across diverse patient populations and compare their performance against established diagnostic standards [12]. Moreover, integrating neural network-based systems into clinical practice requires thorough regulatory approval and validation to ensure patient safety.

While neural networks show immense potential for automated pneumonia detection, several challenges must be addressed. Limited dataset size poses a significant challenge since obtaining medical imaging data can be difficult and expensive. Class imbalance and variability in image quality present additional hurdles. There are also questions about optimally integrating these systems into clinical workflows and ensuring physician trust in model predictions [13]. More research is needed to improve model explainability and transparency for clinicians. Rigorous validation on diverse, real-world populations is essential before clinical deployment [14].

Overall, the application of neural networks to pneumonia detection is an exciting emerging field with many promising avenues. While challenges remain, these innovative techniques could enhance diagnostic accuracy, efficiency, and accessibility. With continued research and validation, neural network-based systems could become invaluable tools for radiologists in improving patient outcomes. However, it is critical that these models are developed and evaluated with a patient-centric approach, ensuring safety, transparency, and equity. Collaborative efforts between medical and technical researchers will be key in realizing the immense promise of this technology.

## 2. Literature Survey

Several studies have explored fundamental CNN architectures for pneumonia detection from chest X-rays. I.e., Jain et al. [15] developed CNN models for pneumonia detection in chest X-rays. They compared a 2-layer and 3-layer CNN, finding the 3-layer achieved an accuracy of 92.31% due to greater depth. They also evaluated pre-trained models VGG16, VGG19, ResNet50, and Inception-v3 on a dataset of 5216 images. VGG19 performed best with 88.46% accuracy. The models focused on maximizing recall to minimize false negatives. The authors conclude that the 3-layer CNN and VGG19 show potential for pneumonia detection, especially with more training data to leverage the pre-trained models.

Other works have customized CNN model designs specifically for pneumonia classification. Bangare et al. [16] developed a customized VGG16 CNN model for pneumonia detection in chest X-rays. Using a dataset of 5863 pediatric images, they progressively increased filter sizes in the convolutional layers. With data augmentation and an Adam optimizer, the model achieved 91.98% test accuracy. Key findings show their modified VGG16 model outperformed other optimizers, indicating it is well-suited for this classification task. They conclude incorporating residual connections, even in shallow CNNs, boosts pneumonia detection performance. The high accuracy demonstrates the potential of custom CNNs for automated pneumonia screening from chest radiographs. Ijaz et al. [17] developed a CNN model for pneumonia detection in chest X-rays. Using a public dataset of 5856 images, they performed resizing, pre-processing, data augmentation, and classification. Augmentation techniques handled class imbalance. Key findings demonstrate combining pre-processing, augmentation, and a customized CNN boosts pneumonia classification performance on chest radiographs. It has been concluded that this approach can assist in automated diagnosis to support physician decision-making.

Some researchers have incorporated segmentation and localization into their models to improve diagnosis. Guendel et al. [18] developed location-aware Dense Networks (DNetLoc) to detect abnormalities in chest X-rays, using the ChestX-Ray14 and PLCO datasets. They incorporated spatial information and high-resolution images, achieving state-of-the-art results on ChestX-Ray14. On PLCO data, DNetLoc improved average AUC by 2.3% for pathologies with location labels. Key findings show location information and high-resolution images boost performance, especially for localized pathologies. The authors propose new patient-wise splits for robust benchmarking on these public datasets. They conclude that exploiting spatial information and image resolution in Dense Networks improves chest x-ray abnormality detection. Alharbi et al. [19] developed an improved BoxENet model using transfer learning from ImgNet and SqueezeNet for pneumonia detection in chest X-rays. They compared performance on full X-rays versus segmented lung images on 4,000 healthy and 3700 pneumonia images. Key findings show the improved BoxENet achieved a decent accuracy for binary and multi-class classification respectively using segmented images. Segmentation improved all models' performance. The improved BoxENet also had faster speeds. In conclusion, incorporating transfer learning in BoxENet with lung segmentation boosts pneumonia classification accuracy and speed compared to standalone CNN models on full chest X-rays.

Advanced training techniques like GAN augmentation and transfer learning have also been applied to enhance pneumonia detection. Srivastav et al. [20] proposed a deep learning model using GAN and transfer learning for pneumonia detection in chest X-rays. A DCGAN was used to generate synthetic images and augment the minority class to balance the dataset. Transfer learning with VGG16 as the base model was then applied for classification. On a dataset of 5856 images, the model achieved 84.5% validation accuracy, outperforming basic CNN and VGG16 without GAN augmentation. Key findings show combining GAN-based oversampling and transfer learning boosts pneumonia classification performance. The authors conclude their approach of generating synthetic images and fine-tuning CNNs enables accurate automated pneumonia screening from chest radiographs. Manickam et al. [21] proposed a deep learning approach using transfer learning with ResNet50, InceptionV3, and InceptionResNetV2 for automated pneumonia detection in chest X-rays. Using a dataset of 5229 images, they applied U-Net-based segmentation and compared optimizer and hyperparameter settings. Key findings show that ResNet50 achieved the best accuracy of 93.06%, outperforming InceptionV3 and InceptionResNetV2. Comparisons demonstrate the proposed models surpass other CNN architectures. The authors conclude transfer learning with fine-tuned ResNet50 enables accurate pneumonia screening from chest radiographs, assisting in clinical diagnosis.

Several studies have compared CNN architectures to identify optimal models for this task. Militante et al. [22] compared six CNN models - AlexNet, GoogleNet, LeNet, VGGNet-16, ResNet-50, and StridedNet - for pneumonia detection in chest x-rays using a dataset of 28,000 images. Models were trained with batch sizes of 32 and 64. The key findings showed that GoogleNet and LeNet achieved the best overall performance, outperforming VGGNet-16. AlexNet and StridedNet also performed well, while ResNet-50 was the poorest performer. In conclusion, GoogleNet and LeNet demonstrated superior performance for pneumonia classification compared to other standard CNN architectures, indicating they are well-suited for this medical imaging task when properly tuned. Jiang et al. [23] proposed an improved VGG16 model called IVGG13 for pneumonia detection in chest X-rays. They reduced the network depth compared to the original VGG16 to avoid overfitting while maintaining feature extraction capacity. Models were trained on 5216 images from a public dataset, with and without data augmentation. Key findings show IVGG13 achieved a higher accuracy of 89.1% versus 74-77% for other CNNs without augmentation and 90% with augmentation. IVGG13 also reduced training time and parameters versus VGG16. The authors conclude that IVGG13 demonstrates superior performance for medical image classification compared to standard CNNs while requiring fewer resources. Szepesi et al. [24] propose a new deep convolutional neural network architecture tailored for automated pneumonia detection in chest X-rays. A key novelty is the use of dropout in the convolutional layers, which improves the model's performance and prevents overfitting. Trained and tested on a dataset of nearly 6,000 pediatric chest X-rays, the model achieves strong results across various evaluation metrics including accuracy, recall, precision, and F1 score. It outperforms networks that rely on transfer learning like VGG, ResNet, and Inception models. Additionally, the model is efficient, making predictions rapidly. Overall, this study shows that a small, customized CNN designed specifically for the task can exceed larger pre-trained models for medical image analysis applications such as pneumonia detection. The results demonstrate the potential of crafting specialized neural network architectures versus relying solely on general pre-trained models.

A few studies have explored Mask R-CNN and SVM-based models. Jaiswal et al. [25] present a deep-learning approach using Mask R-CNN for detecting pneumonia in chest X-rays. Trained on the RSNA pneumonia dataset, their model identifies and localizes lung opacities indicative of pneumonia. To handle class imbalance, they employ aggressive data augmentation. Their model achieves a mean IoU score of 0.199 on the test set, demonstrating its ability to detect pneumonia lesions accurately. Overall, the study shows deep learning can be applied to identify pneumonia in chest X-rays by pinpointing abnormal opacities, helping automate analysis and improve diagnosis. Further improvements in handling class imbalance could enhance performance. Varshini et al. [26] propose a deep-learning model using DenseNet-169 and SVM for pneumonia detection from chest X-rays. DenseNet-169 is employed for feature extraction from the images. Various pre-trained CNNs are evaluated as feature extractors, and

SVM is chosen as the classifier. Hyperparameter optimization further improves performance. On a pneumonia dataset, the proposed model achieves an AUC of 0.80, outperforming prior work. The study demonstrates combining CNN feature extraction with SVM classification can effectively detect pneumonia from chest X-rays. This approach could help automate analysis to assist radiologists, particularly in remote areas.

Finally, a few highly accurate models have been developed using state-of-the-art techniques. Rajpurkar et al. [27] developed a 121-layer convolutional neural network that detects pneumonia from chest X-rays. When tested on 420 X-rays, CheXNet exceeded the average diagnostic performance of 4 radiologists, demonstrating its ability to identify pneumonia at the level of medical experts. The authors also extended CheXNet to detect 14 thoracic diseases, achieving state-of-the-art results on the NIH's ChestX-ray14 dataset. Overall, the study shows deep learning can automate the analysis of chest X-rays to detect pneumonia and other conditions as accurately as radiologists. This could expand access to expert diagnosis where radiologists are limited. Almezhghwi [28] proposes two deep learning methods that combine CNN architectures like AlexNet and VGG16 with support vector machines (SVM) for automated pneumonia detection in chest X-rays. The models use deep networks to extract image features, which are then classified into 12 thoracic diseases by the SVM. When evaluated on the ChestX-ray14 dataset, the combined AlexNet+SVM and VGG16+SVM models achieve improved performance across various metrics compared to using the CNNs alone. The study shows that coupling deep learning for feature extraction with SVM's robust classification capabilities can enhance pneumonia and pathology detection in X-rays. This approach has the potential to help automate the analysis to assist radiologists in evaluating chest radiographs. Overall, the results demonstrate the benefits of blending deep neural networks like CNNs with traditional machine learning techniques like SVM for medical imaging tasks. Chen et al. [29] propose CMT, a deep-learning model combining CNN and Transformer architecture for automated pneumonia detection from ch=est X-rays. CMT employs a convolutional neural network to extract image features and a Transformer module with a novel multi-level, multi-head self-attention mechanism to capture global and local feature relationships. Evaluated on a large COVID-19 chest x-ray dataset, CMT achieves state-of-the-art performance for multi-label classification of pneumonia with 99.7% accuracy for COVID-19. The self-attention module is more efficient, requiring less training and inference time than standard attention. Overall, the study demonstrates CMT's ability to recognize pneumonia from X-rays while being interpretable and rapid accurately.

In the past, very few studies concentrated on Transformers and ensembling models for pneumonia detection from chest X-rays, with most works using standalone CNNs or Transformers. Our work demonstrates a novel fusion model combining ResNet and Vision Transformer networks via model-level ensembling. This showcases the potential of blending CNN and Transformer models to harness their complementary strengths, which has not been explored in prior works. Our ensemble approach achieves state-of-the-art performance, proving that fusing CNN and Transformer models can boost pneumonia classification accuracy compared to using individual architectures. While most previous studies focused on solitary models, our experiments reveal the value of novel ensemble techniques to advance the state of the art in this domain.

## 3. Methods

### 3.1 Datasets

In this work, the Kaggle pneumonia dataset [30] is created using chest X-ray images of pediatric patients aged 1-5 selected from Guangzhou Women and Children's Medical Center's retrospective cohort's dataset samples are provided in Figure 1. The X-rays were taken as part of routine clinical care, and the dataset is labeled into two classes: normal and Pneumonia. The dataset used for training includes 1341 normal samples and 3875 pneumonia samples. The test set contains 234 normal samples and 390 pneumonia samples. Due to its small size of only 16 images, the validation dataset provided by Kaggle was not utilized for experimentation. Therefore, the training dataset was split into train and validation datasets based on classes, with an 80:20 ratio, and detailed information can be found in Table 1.

| Set | Normal | Pneumonia | Total |
|---|---|---|---|
| Train | 1060 | 3112 | 4172 |
| Val | 281 | 763 | 1044 |
| Test | 234 | 390 | 625 |

Table 1. Dataset samples for each set

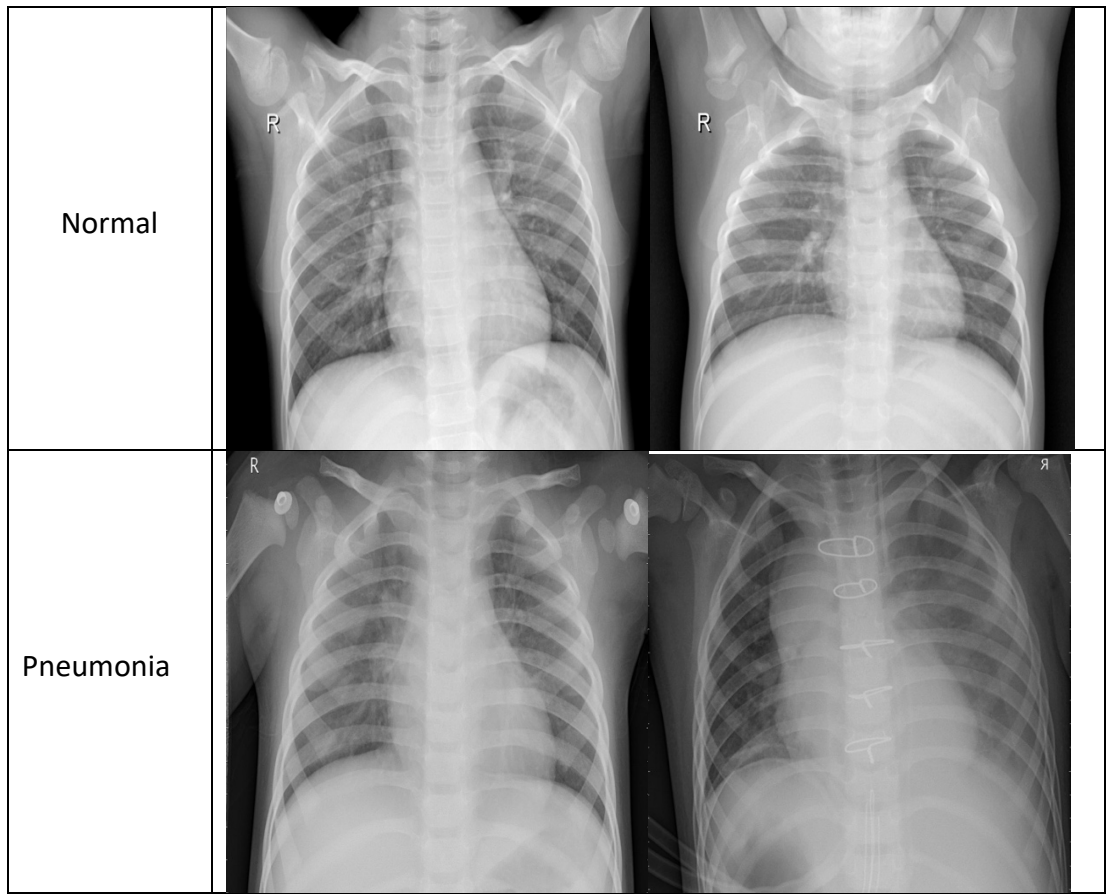

Figure 1. The first row displays the Normal chest X-rays, and the second row shows Pneumonia samples.

**3.2 Description of Models**

A fusion network combining CNN and Transformers is designed to run the experiments. The fusion model is created using the ResNet-34 variant and multi-Axis Vision transform small variant model.

**ResNet [31] :**

ResNet, a convolutional neural network architecture introduced in 2015, is short for "Residual Network." It has been a groundbreaking innovation in computer vision and deep learning. It has greatly impacted image-related tasks such as image classification, object detection, and image segmentation. The development of ResNet addressed the problem of vanishing gradients in very deep neural networks. As the depth of a neural network increases, the gradient signal diminishes during backpropagation, making training increasingly difficult. The key insight of ResNet is the introduction of skip connections, which allow for easier gradient flow through the network. The core building block of ResNet is the "residual block" or "identity block," which contains two or three convolutional layers with shortcut connections that bypass one or more layers in the block. These shortcut connections, also known as "identity" or "shortcut" connections, provide a direct path for information to flow through the network, helping to mitigate the vanishing gradient problem and allowing for the training of very deep networks. ResNet 34 variant is used for experiments and the detailed composition of the model is in Figure 2.

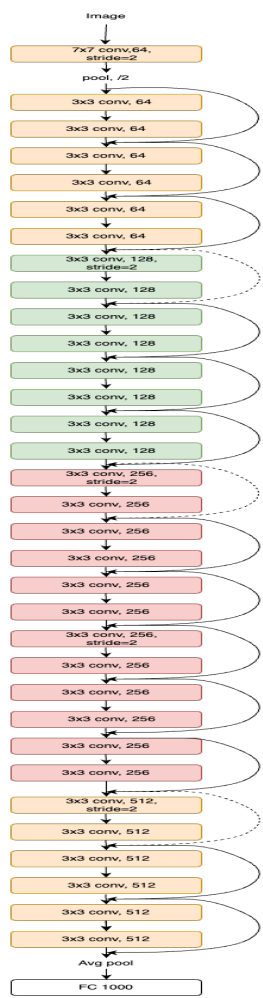

Figure 2. ResNet34 model architecture

**Multi-Axis Vision Transformer [32]:**
Recently, Transformers have gained attention in computer vision. However, their self-attention mechanisms' scalability with image size has limited their adoption in state-of-the-art vision backbones. Multi-Axis Vision Transformer (MaxVit) solves scalability issues by stacking repeated blocks of Max-SA, modules, and MBConv hierarchically [cite this one]. Max-SA is a fundamental architecture component that includes blocked local and dilated global attention, providing global perception in linear complexity. It can carry out both local and global spatial interactions within a single block. As compared to full self-attention, Max-SA is highly capable. We show the detailed composition of MaxViT in Figure 3.

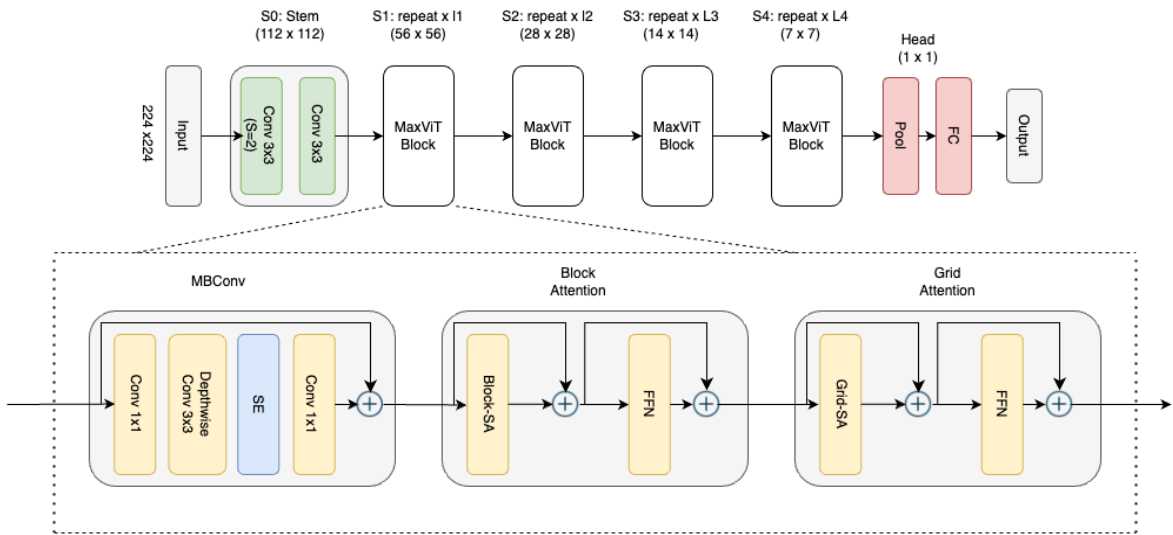

Figure 3. MaxViT model architecture

The fusion model comprises two different paths, as shown in the diagram. The first path is the ResNet34 model, while the second path is the MaxViT base model. Both models are initialized with pre-trained weights from ImageNet. The output layer of each model is removed, and the features from both paths are combined using a flattening layer. Finally, the combined features are passed through a dense layer with two nodes to generate prediction probabilities. Figure 4. displays a comprehensive view of the architecture.

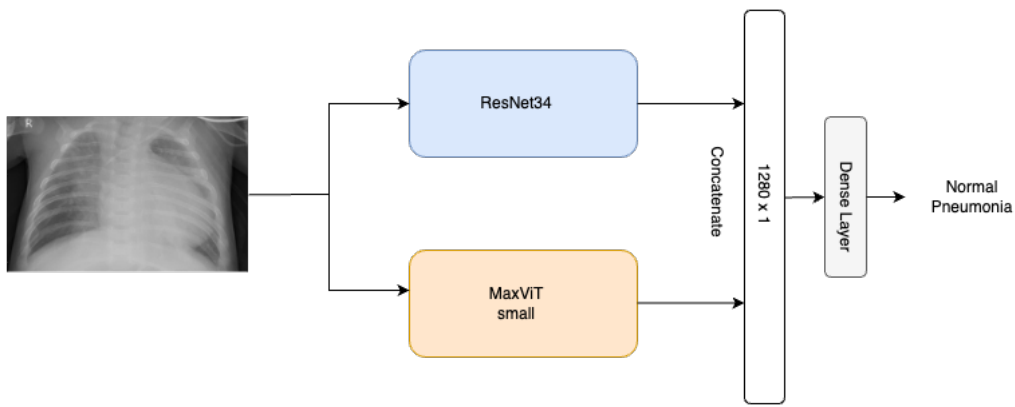

Figure 4. Proposed model architecture.

## 4. Experimental setup:

All the models are trained using transfer learning and are initialized with pre-trained weights on ImageNet. The models have been implemented in PyTorch [33] using the timm library [34] and trained for 100 epochs with warmup epochs of 5 using the Adam optimizer [35] with a momentum of 0.9. A mini-batch size of 8 has been used for training, with a weight decay of 5 x $10^{-2}$ by default. The learning rate has been initialized to 5 x $10^{-5}$ and decreases with the cosine schedule. Data augmentation techniques such as random cropping, horizontal flip, vertical flip, and color have been used during training. The experiments are conducted on NVIDIA RTX 3090. The best model during training is selected based on accuracy.

## 5. Performance criteria:

To assess our trained model's performance effectively, we use various widely accepted and reliable metrics in the medical field. The performance of the models are evaluated using the following metrics 1) Accuracy 2) Kappa score 3) Sensitivity 4) Specificity  5) PPV (Positive Predictive Value). These metrics are expressed in **Eqs 1-5**

*Accuracy = (TP + TN)/(TP + FP + TN + FN)*
*Kappa score = $(P_o - P_e)/(1 - P_e)$*
*Sensitivity = TP / (TP + FN)*
*Specificity = TN / (TN + FP)*
*PPV = TP / (TP + FP)*

"TP" represents the number of true positive samples in a category, while "FN" means the number of false negative samples. Similarly, "TN" denotes the number of true negative samples, and "FP" indicates the number of false positive samples in a category.

| Model | Accuracy | Kappa score | Sensitivity | Specificity | PPV |
|---|---|---|---|---|---|
| ResNet34 [31] | 0.9407 | 0.8696 | 0.9974 | 0.8461 | 0.9152 |
| vit_tiny_patch16 [36] | 0.9407 | 0.8698 | 0.9948 | 0.8504 | 0.9172 |
| swin_tiny_patch4_window7 [37] | 0.9455 | 0.8804 | 0.9974 | 0.8589 | 0.9218 |
| **Resnet34 + maxvit_small (Ours)** | **0.9487** | **0.8875** | **1.0** | **0.8632** | **0.92417** |

Table. 1 Experimental comparison results between our model and other SOTA models for Normal vs. Pneumonia classification

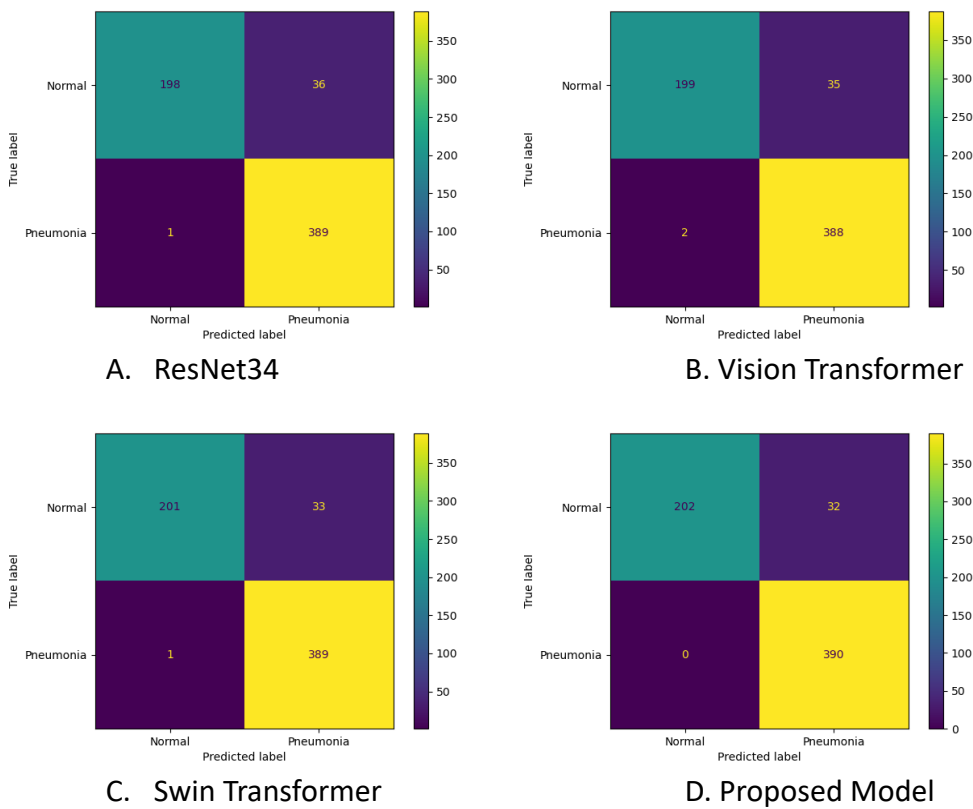

A. ResNet34  
B. Vision Transformer  
C. Swin Transformer  
D. Proposed Model  

Figure 4. Confusion matrix of models on Pneumonia detection

## 6. Results and Analysis:

We calculated accuracy, kappa, sensitivity, specificity, and positive predictive value to evaluate performance. Our model was compared to several CNN and Vision Transformer models, including Resnet34, Vision Transformer, and Swin-transformer tiny variant, all trained under the same settings. Our model achieved a SOTA level of 94.87% accuracy for classification, as shown in Table 2. Similarly, our model achieves a kappa score of 0.88 and a sensitivity of 1.0. Additionally, our model outperformed in other metrics, as presented in Table 2. Analyzing the confusion matrix for the test set in Figure 4, indicates that our suggested architecture has fewer misclassifications than the other existing models. Our model demonstrated good generalization performance on the test dataset by combining the resnet34 and MaxVit variants via model-level ensembling. This combination of CNN and Vision Transformer models allows for learning robust features for pneumonia classification.

## 7. Conclusion

Our research paper proposed a novel model-level ensembled architecture for detecting pneumonia using chest X-rays. To achieve this, we utilized a combination of the resnet34 and MaxVit small variants models, which helped us create a highly effective model for identifying pneumonia. We experimentally demonstrate the efficacy of the proposed model over CNN or pure vision transformer-based models by achieving SOTA performance as measured by various classification metrics. Moreover, our model outperformed other popular neural networks and transformer models, which is a promising outcome for the future of pneumonia detection. Looking ahead, we intend to continue exploring various combinations of architectures and datasets further to enhance the accuracy and reliability of our model.

Overall, Our work reveals promising new research directions in employing model ensembling to combine diverse strengths for improved pneumonia detection, which has immense real-world implications. With further development, such systems could be deployed in clinical settings to provide radiologists with accurate decision support and enhanced productivity. We aim to extend this research by exploring different ensemble configurations, larger datasets, and real-world validation studies. Overall, we successfully showcase the potential of fusing CNNs and Vision Transformers via model-level ensembling to attain new state-of-the-art performance for automated pneumonia diagnosis using chest X-rays.

**Acknowledgment**

This study was made possible thanks to the financial support of Tror(tror.ai). We deeply appreciate their support, which enabled us to conduct the experiments, analyze the data, and disseminate the findings.

**References:**

[1] K. K. Yadav and S. Awasthi, "Childhood Pneumonia: What's Unchanged, and What's New?," *Indian Journal of Pediatrics*, vol. 90, no. 7. 2023. doi: 10.1007/s12098-023-04628-3.

[2] C. L. Fischer Walker *et al.*, "Global burden of childhood pneumonia and diarrhoea," *The Lancet*, vol. 381, no. 9875. 2013. doi: 10.1016/S0140-6736(13)60222-6.

[3] H. J. Zar, S. A. Madhi, S. J. Aston, and S. B. Gordon, "Pneumonia in low and middle income countries: Progress and challenges," *Thorax*, vol. 68, no. 11. 2013. doi: 10.1136/thoraxjnl-2013-204247.

[4] P. Gang *et al.*, "Dimensionality reduction in deep learning for chest X-ray analysis of lung cancer," in *Proceedings - 2018 10th International Conference on Advanced Computational Intelligence, ICACI 2018*, 2018. doi: 10.1109/ICACI.2018.8377579.

[5] N. Barakat, M. Awad, and B. A. Abu-Nabah, "A machine learning approach on chest X-rays for pediatric pneumonia detection," *Digit Health*, vol. 9, 2023, doi: 10.1177/20552076231180008.

[6] H. Sharma, J. S. Jain, P. Bansal, and S. Gupta, "Feature extraction and classification of chest X-ray images using CNN to detect pneumonia," in *Proceedings of the Confluence 2020 - 10th International Conference on Cloud Computing, Data Science and Engineering*, 2020. doi: 10.1109/Confluence47617.2020.9057809.

[7] J. G. Lee *et al.*, "Deep learning in medical imaging: General overview," *Korean Journal of Radiology*, vol. 18, no. 4. 2017. doi: 10.3348/kjr.2017.18.4.570.

[8] A. Krizhevsky, I. Sutskever, and G. E. Hinton, "ImageNet classification with deep convolutional neural networks," *Commun ACM*, vol. 60, no. 6, 2017, doi: 10.1145/3065386.

[9] Y. LeCun, G. Hinton, and Y. Bengio, "Deep learning (2015), Y. LeCun, Y. Bengio and G. Hinton," *Nature*, vol. 521, 2015.

[10] A. Esteva *et al.*, "A guide to deep learning in healthcare," *Nature Medicine*, vol. 25, no. 1. 2019. doi: 10.1038/s41591-018-0316-z.

[11] L. Oakden-Rayner, J. Dunnmon, G. Carneiro, and C. Re, "Hidden stratification causes clinically meaningful failures in machine learning for medical imaging," in *ACM CHIL 2020 - Proceedings of the 2020 ACM Conference on Health, Inference, and Learning*, 2020. doi: 10.1145/3368555.3384468.

[12] M. Roberts *et al.*, "Common pitfalls and recommendations for using machine learning to detect and prognosticate for COVID-19 using chest radiographs and CT scans," *Nat Mach Intell*, vol. 3, no. 3, 2021, doi: 10.1038/s42256-021-00307-0.

[13] F. Jiang *et al.*, "Artificial intelligence in healthcare: Past, present and future," *Stroke and Vascular Neurology*, vol. 2, no. 4. 2017. doi: 10.1136/svn-2017-000101.

[14] J. R. Zech, M. A. Badgeley, M. Liu, A. B. Costa, J. J. Titano, and E. K. Oermann, "Variable generalization performance of a deep learning model to detect pneumonia in chest radiographs: A cross-sectional study," *PLoS Med*, vol. 15, no. 11, 2018, doi: 10.1371/journal.pmed.1002683.

[15] R. Jain, P. Nagrath, G. Kataria, V. Sirish Kaushik, and D. Jude Hemanth, "Pneumonia detection in chest X-ray images using convolutional neural networks and transfer learning," *Measurement (Lond)*, vol. 165, 2020, doi: 10.1016/j.measurement.2020.108046.

[16] Dr. Sunil L. Bangare, Hrushikesh S. Rajankar, Pavan S. Patil, Karan V. Nakum, and Gopal S. Paraskar, "Pneumonia Detection and Classification using CNN and VGG16," *International Journal of Advanced Research in Science, Communication and Technology*, 2022, doi: 10.48175/ijarsct-3851.

[17] A. Ijaz, S. Akbar, B. Alghofaily, S. A. Hassan, and T. Saba, "Deep Learning for Pneumonia Diagnosis Using CXR Images," in *Proceedings - 2023 6th International Conference of Women in Data Science at Prince Sultan University, WiDS-PSU 2023*, 2023. doi: 10.1109/WiDS-PSU57071.2023.00023.

[18] S. Gündel, S. Grbic, B. Georgescu, S. Liu, A. Maier, and D. Comaniciu, "Learning to recognize abnormalities in chest X-rays with location-aware dense networks," in *Lecture Notes in Computer Science (including subseries Lecture Notes in Artificial Intelligence and Lecture Notes in Bioinformatics)*, 2019. doi: 10.1007/978-3-030-13469-3_88.

[19] A. H. Alharbi and H. A. Hosni Mahmoud, "Pneumonia Transfer Learning Deep Learning Model from Segmented X-rays," *Healthcare (Switzerland)*, vol. 10, no. 6, 2022, doi: 10.3390/healthcare10060987.

[20] D. Srivastav, A. Bajpai, and P. Srivastava, "Improved classification for pneumonia detection using transfer learning with GAN based synthetic image augmentation," in *Proceedings of the Confluence 2021: 11th International Conference on Cloud Computing, Data Science and Engineering*, 2021. doi: 10.1109/Confluence51648.2021.9377062.

[21] A. Manickam, J. Jiang, Y. Zhou, A. Sagar, R. Soundrapandiyan, and R. Dinesh Jackson Samuel, "Automated pneumonia detection on chest X-ray images: A deep learning approach with different optimizers and transfer learning architectures," *Measurement (Lond)*, vol. 184, 2021, doi: 10.1016/j.measurement.2021.109953.

[22] S. V. Militante, N. V. Dionisio, and B. G. Sibbaluca, "Pneumonia and COVID-19 Detection using Convolutional Neural Networks," in *Proceeding - 2020 3rd International Conference on Vocational Education and Electrical Engineering: Strengthening the framework of Society 5.0 through Innovations in Education, Electrical, Engineering and Informatics Engineering, ICVEE 2020*, 2020. doi: 10.1109/ICVEE50212.2020.9243290.

[23] Z. P. Jiang, Y. Y. Liu, Z. E. Shao, and K. W. Huang, "An improved VGG16 model for pneumonia image classification," *Applied Sciences (Switzerland)*, vol. 11, no. 23, 2021, doi: 10.3390/app112311185.

[24] P. Szepesi and L. Szilágyi, "Detection of pneumonia using convolutional neural networks and deep learning," *Biocybern Biomed Eng*, vol. 42, no. 3, 2022, doi: 10.1016/j.bbe.2022.08.001.

[25] A. K. Jaiswal, P. Tiwari, S. Kumar, D. Gupta, A. Khanna, and J. J. P. C. Rodrigues, "Identifying pneumonia in chest X-rays: A deep learning approach," *Measurement (Lond)*, vol. 145, 2019, doi: 10.1016/j.measurement.2019.05.076.

[26] D. Varshni, K. Thakral, L. Agarwal, R. Nijhawan, and A. Mittal, "Pneumonia Detection Using CNN based Feature Extraction," in *Proceedings of 2019 3rd IEEE International*

[26] *Conference on Electrical, Computer and Communication Technologies, ICECCT 2019*, 2019. doi: 10.1109/ICECCT.2019.8869364.

[27] P. Rajpurkar *et al.*, "CheXNet: Radiologist-Level Pneumonia Detection on Chest X-Rays with Deep Learning." [Online]. Available: https://stanfordmlgroup.

[28] K. Almezhghwi, S. Serte, and F. Al-Turjman, "Convolutional neural networks for the classification of chest X-rays in the IoT era," *Multimed Tools Appl*, vol. 80, no. 19, 2021, doi: 10.1007/s11042-021-10907-y.

[29] S. Chen, S. Ren, G. Wang, M. Huang, and C. Xue, "Interpretable CNN-Multilevel Attention Transformer for Rapid Recognition of Pneumonia from Chest X-Ray Images," *IEEE J Biomed Health Inform*, 2023, doi: 10.1109/JBHI.2023.3247949.

[30] D. S. Kermany *et al.*, "Identifying Medical Diagnoses and Treatable Diseases by Image-Based Deep Learning," *Cell*, vol. 172, no. 5, 2018, doi: 10.1016/j.cell.2018.02.010.

[31] K. He, X. Zhang, S. Ren, and J. Sun, "Deep residual learning for image recognition," in *Proceedings of the IEEE Computer Society Conference on Computer Vision and Pattern Recognition*, 2016. doi: 10.1109/CVPR.2016.90.

[32] Z. Tu *et al.*, "MaxViT: Multi-Axis Vision Transformer," Apr. 2022, Accessed: May 18, 2023. [Online]. Available: http://arxiv.org/abs/2204.01697

[33] A. Paszke *et al.*, "PyTorch: An imperative style, high-performance deep learning library," in *Advances in Neural Information Processing Systems*, 2019.

[34] R. Wightman *et al.*, "rwightman/pytorch-image-models: v0.8.10dev0 Release," Feb. 2023, doi: 10.5281/ZENODO.7618837.

[35] D. P. Kingma and J. Ba, "Adam: A Method for Stochastic Optimization," Dec. 2014, Accessed: May 02, 2023. [Online]. Available: http://arxiv.org/abs/1412.6980

[36] A. Dosovitskiy *et al.*, "An Image is Worth 16x16 Words: Transformers for Image Recognition at Scale," Oct. 2020, Accessed: May 01, 2023. [Online]. Available: http://arxiv.org/abs/2010.11929

[37] Z. Liu *et al.*, "Swin Transformer: Hierarchical Vision Transformer using Shifted Windows," Mar. 2021, Accessed: May 01, 2023. [Online]. Available: http://arxiv.org/abs/2103.14030